# A Random Sequential Addition-based Algorithm for Generating Random Packings of Non-spherical Particles with Tunable Topology

Amirmehdi Salehi, Gholamreza Pircheraghi[*]

**Abstract**

Despite many advantages associated with the use of spherical particles, in practice, non-spherical particles are the main ingredient in sintering-based processing techniques, like selective-laser-sintering, where their shape and surface roughness influence the formation and development of inter-particle contacts. Not only do particle-particle contacts influence the overall pore surface curvature and thus the driving force behind the sintering process, but they also affect the development of macroscopic properties, especially the mechanical strength. To study the role of particle connectivity in the sintering process, we propose a simple algorithm that generates random packings of non-spherical particles in a wide range of particle connectivities. We create irregular particle shapes by clustering three groups of spheres in three separate stages that allow us to tune the shape and surface roughness of the particles. Packings generated by our algorithm can be used both in simulations and for printing prototypes that can be tested in real-time experiments.

## 1 Introduction

In general, there is a tendency for using spherical particles in sintering-based processing techniques as they reduce the challenges involved in packing, flowability [1, 2], and heat conduction and facilitate numerical modeling [3, 4]. However, obtaining spherical particles is not always an easy task, and sometimes their cost is not financially justifiable. Therefore, many sintering-based techniques, such as selective laser sintering (SLS), are generally carried out using non-spherical particles [5].

Despite the use of irregularly-shaped particles in practice, theoretical groundwork and modeling are often based on the assumption of spherical particles. For instance, the pioneering viscous sintering model proposed by Frenkel [6] and the majority of later models on the kinetics of sintering were all based on the assumption of spherical particles [7]. However, these types of simplifying presuppositions can lead to inconsistent and false conclusions, especially when the objective is to study the connectivity of the particles. This possible inconsistency is because the inter-particle connectivity is strongly affected by the shape and surface roughness of the particles, which are often overlooked in the study of sintering processes. The development of inter-particle contacts influences the sintering kinetics and plays a significant role in the evolution of macroscopic properties in sintered parts [8-11]. For instance, we observed in another work, that the macroscopic permeability of a sintered specimen decreases with the formation of interparticle contacts [11]. These contacts were also observed to help partition the irregularly large pore regions in the sintered specimens and thus modify the pore size distribution. We also noted the interdependent role of chain diffusion and particle-particle contacts in the development of compressive modulus in sintered porous parts [11]. Therefore, there should be an interest in taking the development of inter-particle contacts into consideration when modeling or studying sintering-based processes.

In this regard, despite the ever-improving capabilities in 3D-reconstruction techniques and the increasing computational power of computers, reliable detection of interfacial contacts in 3D images, especially for non-spherical particles, is not a trivial task [12]. Therefore, a viable approach to analyzing the role of inter-particle contacts is to generate random packings of non-spherical particles with tunable shape and surface roughness that can capture various topologies and connectivities. These packings can be used in computer simulations and printing 3D prototypes suitable for experimental testing. In this regard, there have been many attempts at generating random particle packings by modeling forces at particle level [13-15]. However, modeling such forces for irregularly shaped particles is a non-trivial task and is even made more challenging if the precise connectivity of the particles is desired; this renders these simulations computationally complex. Here, we propose a simple, lightweight algorithm that produces packings of non-spherical particles with tunable topology and connectivity based on a random sequential addition (RSA) algorithm.

## 2 Algorithm

### 2.1 Background

To generate a non-spherical particle, we assemble overlapping varying-sized spheres in a cluster. This approach for the construction of irregularly-shaped particles affords the ability to tune the shape and surface roughness of the particles using an efficient algorithm.

Figure 1 shows the optical and SEM micrographs of a few common polymer powders used in sintering-based processes. These are examples of real particle shapes that can be suitably reproduced by assembling varying-sized spheres. For instance, Figure 2 schematically demonstrates how, in theory, one can adequately capture the shape and surface roughness of non-spherical particles using clusters of overlapping spheres.

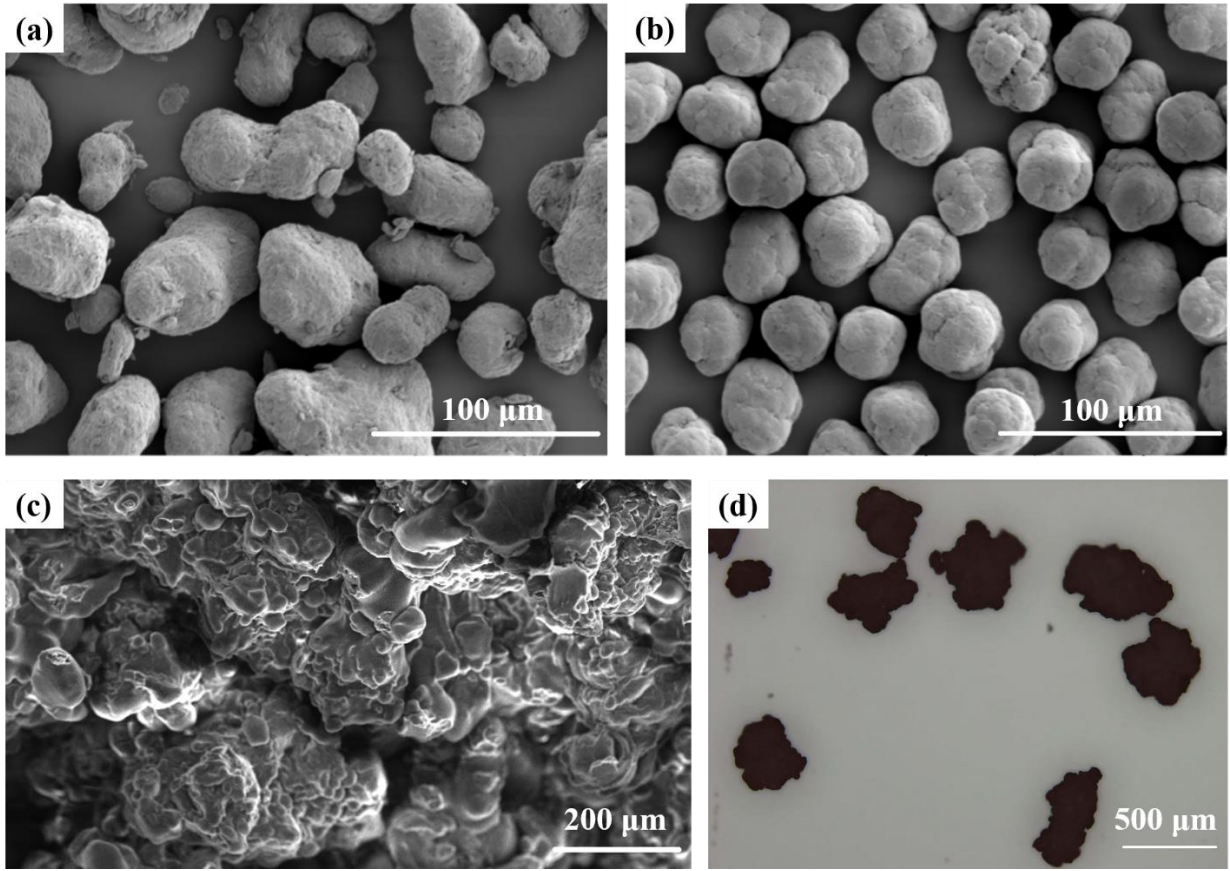

Figure 1. Micrographs of (a) PA12 particles produced by precipitation, (b) PA12 manufactured by direct polymerization [16], (c, d) High-density Polyethylene (HDPE) used in sintering processes [11].

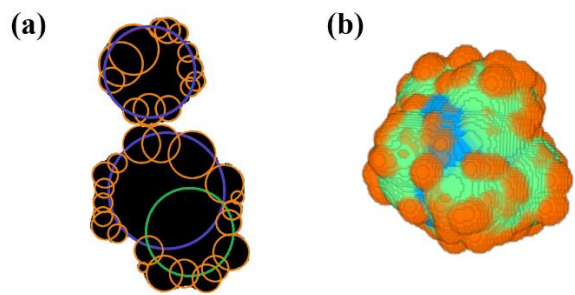

Figure 2. (a) Representation of the outline of a real particle with overlapping varying-sized circles. (b) The three layers of spheres clustered to form a non-spherical particle. The blue sphere is the core, green spheres are the secondary spheres, and the orange ones are the tertiary spheres.

**2.2 Reproducing non-spherical particles**

Generally, each particle is formed by clustering spheres in three different stages. In the first stage, the particle core is established; specifically, for each particle, one sphere is seeded in a representative volume element (RVE) by random sequential addition (RSA). The spheres seeded in this stage are non-overlapping. After establishing the particle cores, we form their overall shape by adding a second layer of spheres onto the existing cores. Finally, by attaching yet another layer of smaller spheres, we tune the roughness of the particles' surface. The majority of inter-particle bonds are formed during the final stage. Figure 2 shows an example of three layers of spheres assembled to form a non-spherical particle.

To efficiently reproduce non-spherical particles, our algorithm should be able to cluster the constituent spheres of the particles according to two conditions. First, spheres have to be tightly packed, so the resultant particles are bulk with no intra-particle pores (Condition 1). Second, adding any sphere onto a particle should always lead to the growth of that particle (Condition 2).

Condition 1: To ensure tight packing, each newly added sphere should intersect at least another existing sphere with their center-to-center distance satisfying the following condition:

$$d \leq R_2 - (\alpha \times R_1) \tag{1}$$

where $d$ is the center-to-center distance for the two spheres, $R_1$ and $R_2$ are the radius of the two spheres ($R_2 > R_1$), and $0 \leq \alpha \leq 1$ is the compactness factor.

As shown in Figure 3, the lower the compaction factor, the more loosely the particle's constituent spheres will be packed, and the more likely that intra-particle holes or voids will exist in particles. It should be realized, however, that despite giving rise to bulk particles, a high compactness factor means that for the same volume, a single particle is made up of a larger number of spheres, which will degrade the performance of the algorithm. In this regard, based on our assessments for a wide range of particle sizes, a compactness factor of 0.5 is a good trade-off.

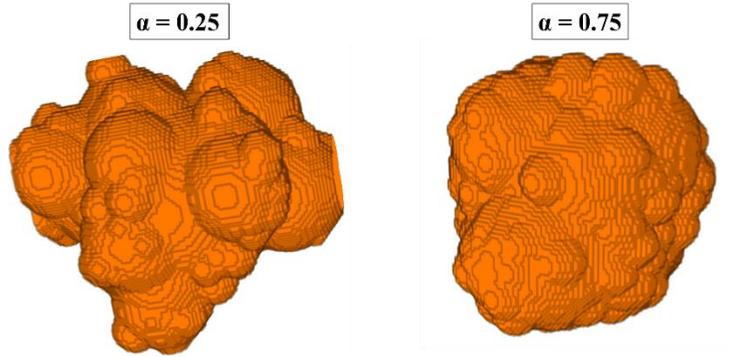

Figure 3. Two particles produced with two different compactness factors.

Condition 2: To ensure efficient sphere clustering, no newly added sphere should be wholly contained within any of the existing spheres. In other words,

$$d > R_2 - R_1 \qquad (2)$$

This condition will improve the performance of the algorithm as it ensures that each added sphere will inevitably lead to the growth of the particles.

**2.3 Tracking the connectivity of particles**

To be able to monitor the formation of inter-particle contacts, we have to categorize each newly added sphere into one of two groups: interfacial or bulk. Interfacial spheres are shared by two or more adjacent particles, and bulk spheres belong to only one particle. To determine the type of a newly-added sphere, we have to find all the spheres that overlap it and identify the particles they belong to. In this regard, a simplistic approach is to calculate the center-to-center distance of the new sphere with its neighbors and check if the following condition holds for any of them.

$$d < R_2 + R_1 \tag{3}$$

However, due to the pixelated nature of the spheres, such calculations will not deliver reliable results. For instance, each generated sphere surface voxel in our algorithm is located up to 0.5 away from a real sphere with the same center and radius in the floating numbers' domain. Therefore, to detect the overlapping spheres reliably, instead of using the center-to-center distance, we scan the enclosing pixels of the newly-added sphere. In other words, we detect overlap by pixel-adjacency inspection. It should be pointed that we regard two spheres as overlapping if at least two of their surface voxels are 1-connected or 2-connected. 0, 1, and 2-connectedness of voxels are demonstrated in Figure 4.

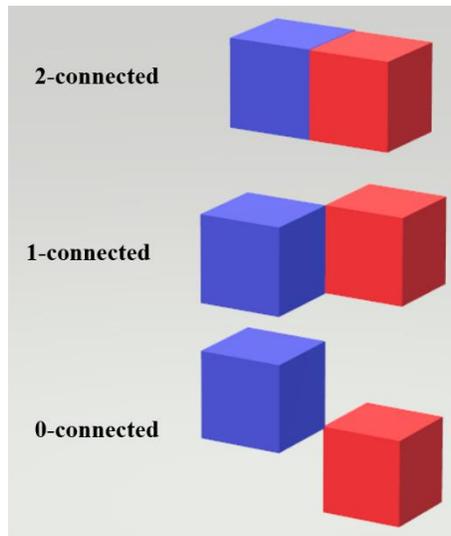

Figure 4. Schematic illustration of the 0, 1, and 2-connectedness of the adjacent voxels.

Moreover, to have robust control over the connectivity of the particles, we have to make sure that (1) each particle-particle contact occurs through one interfacial sphere only, and (2) each interfacial sphere accounts for one and only one connection. For the first condition, we preclude any two interfacial spheres from overlapping, whereas for the second condition, we have to make sure that all the spheres, be it interfacial or bulk, are tightly packed, so no unwanted holes are left in their clusters (see section(2.2)). Precluding unwanted holes is vital because, if left unchecked, they would lead to unreliable results when it comes to monitoring the exact number of particle-particle contacts. Figure 5 illustrates the overall procedure of adding secondary and tertiary spheres onto the core spheres.

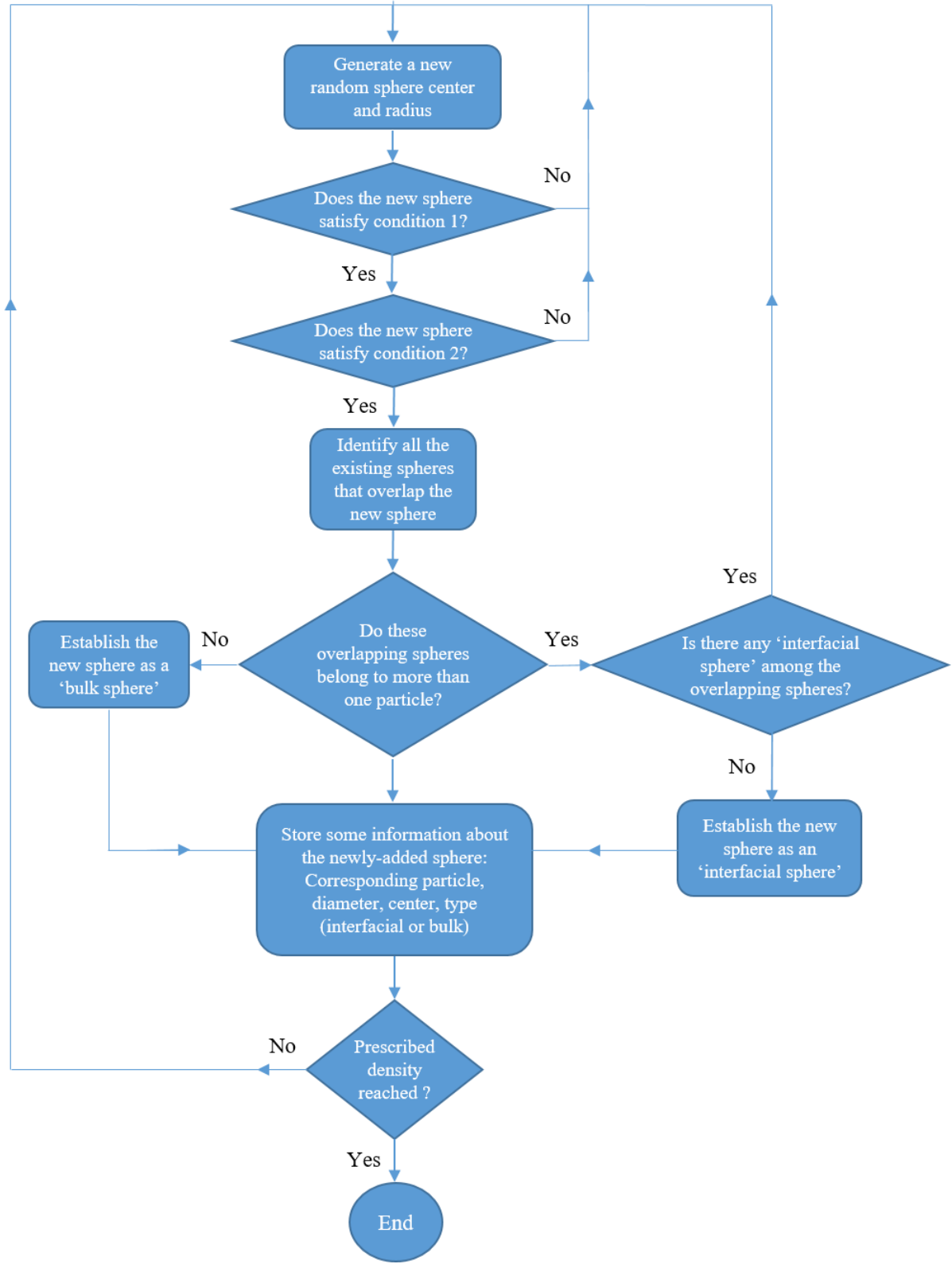

Figure 5. A schematic of the flow of the program for adding secondary and tertiary spheres.

**2.4 Input Parameters**

The inputs of the algorithm are the core, secondary, and tertiary sphere radii ($r_1$, $r_2$, and $r_3$), the volume fraction of tertiary spheres (X), and the final packing density. The effects of tuning these parameters are elaborated in the following subsections.

**2.4.1 Core radius ($r_1$)**

The strategy of generating non-spherical particles by clustering three layers of varying-sized overlapping spheres affords the ability to control the final particle size distribution in packings. In other words, as will be shown in the following sections, the secondary and tertiary spheres mainly control the shape and roughness of the particles without increasing the core radius. This is a useful feature that affords control over the size distribution of the constituent particles. Particle sizes in packings generated by our algorithm are at most within two voxels of the predetermined core radii. Therefore, the appropriate core radius in the first stage is the desired final particle radius. For instance, when trying to generate packings from real particles, it is possible to use the equivalent circle or sphere radius of the particles obtained from their real 2D or 3D images (see Figure 1) as the core radius.

**2.4.2 Secondary and tertiary sphere radii ($r_2$ and $r_3$)**

As discussed in the previous section, the radius of the spheres in the second and third stage controls the shape and surface roughness of the particles and thus their connectivity in the generated random packings.

To show this relationship, we introduce some factors that quantify the shape and connectivity of the particles. In this regard, we make use of an index called the sphericity index (SI) to quantify the shape of the constituent particles in packings. The SI lies between 0 (non-spherical) and 1 (spherical) and is defined as the following [17]:

$$SI = 36\pi \frac{v^2}{s^3} \tag{4}$$

where $v$ and $s$ are the volume and surface area of the particles, respectively. Also, the packings' connectivity is tracked by monitoring the mean coordination number ($CN = \text{total number of contacts}/\text{number of particles}$). Table 1 and Table 2 show the variation of the SI and CN of the particles with $r_2$ and $r_3$, respectively.

Table 1. the variation of the SI and CN with $r_2$.
$r_1$ = 30-40, $r_3$ = 10-15, density = 0.65, RVE = $300^3$ voxels, number of particles = 88 ± 3. (All the radii are chosen randomly from a uniform distribution in the ranges shown.)

| $r_2$ | SI | CN | mean particle radius |
|---|---|---|---|
| 25-35 | 0.68 ± 0.01 | 3.87 ± 0.08 | 35.63 ± 0.29 |
| 20-30 | 0.67 ± 0.00 | 4.00 ± 0.06 | 35.27 ± 0.28 |
| 15-25 | 0.64 ± 0.01 | 4.40 ± 0.11 | 35.59 ± 0.34 |

Table 2. the variation of the SI and mean coordination number of the particles with $r_3$.
$r_1$ = 30-40, $r_2$ = 20-30, density = 0.65, RVE = $300^3$ voxels, number of particles = 88 ± 3.

| $r_3$ | SI | CN | mean particle radius |
|---|---|---|---|
| 15-20 | 0.70 ± 0.00 | 3.75 ± 0.05 | 35.11 ± 0.48 |
| 10-15 | 0.67 ± 0.00 | 4.00 ± 0.06 | 35.27 ± 0.28 |
| 5-10 | 0.59 ± 0.00 | 4.35 ± 0.02 | 34.89 ± 0.25 |

As evident from the above tables, by decreasing the size of the secondary and tertiary spheres, we can reduce the sphericity of the particles and obtain packings with a higher mean coordination number. It should be noted that the effect of the third stage spheres on the sphericity of the particles is more pronounced compared to secondary spheres. Figure 6 demonstrates how changing the radii of the particles in the third stage affects the surface roughness of the generated particles and thus the topology of the constructed packings.

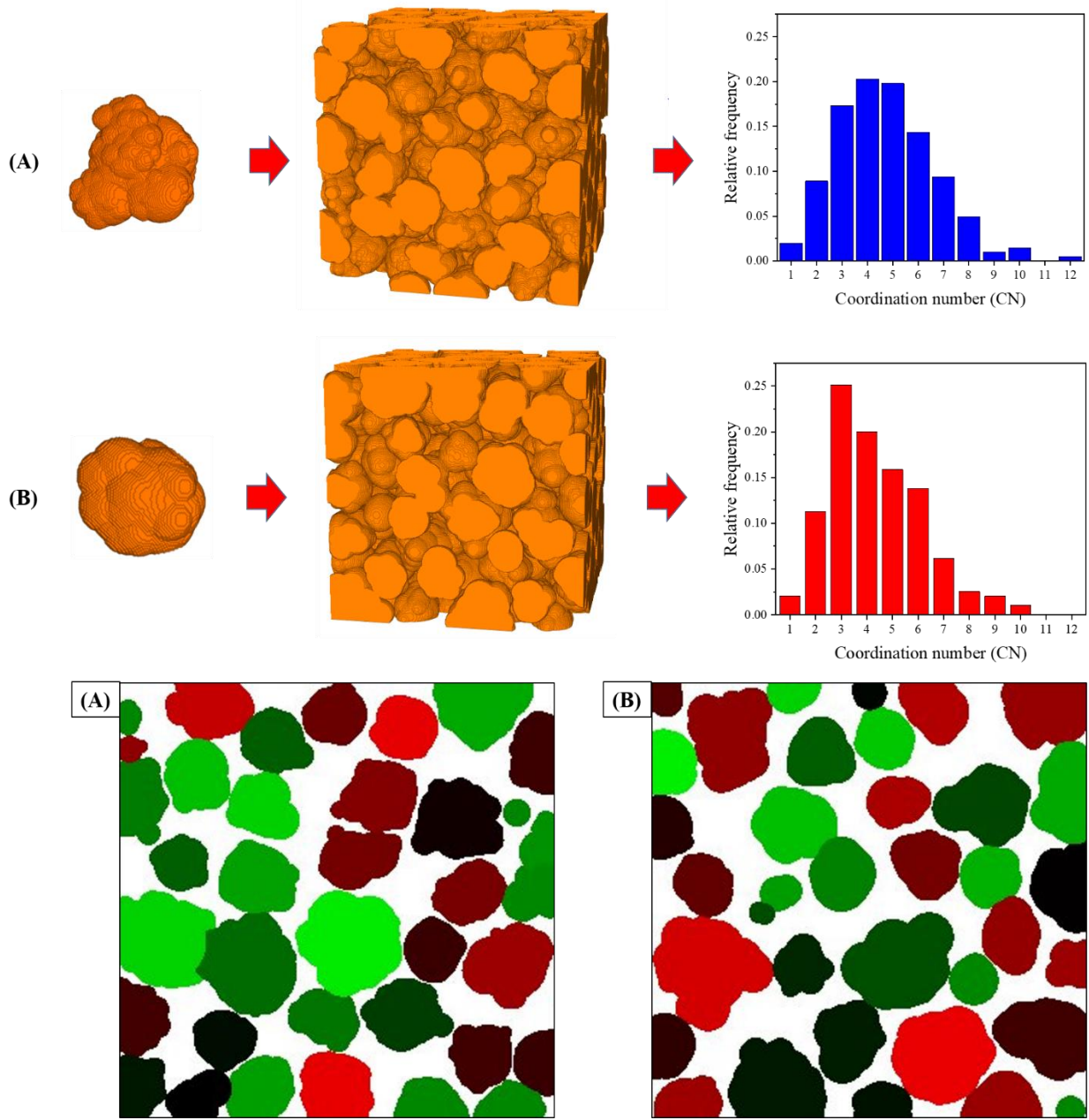

Figure 6. Two random packings with different size ranges for tertiary spheres, (a) $5 < r_3 < 10$, (b) $10 < r_3 < 15$. Other inputs are: $20 < r_1 < 30$, $15 < r_2 < 20$, density = 0.65, and RVE=$300^3$ voxels. The two 2D images on the bottom are cross-sections of the above two packings, which demonstrate the surface roughness of the particles.

### 2.4.3 Volume fraction of tertiary spheres (x)

Another parameter that can be used to tune the surface roughness, shape, and connectivity of the particles is the percent contribution of the tertiary spheres (x). Table 3 shows that by decreasing x, the SI and CN rise and fall, respectively. In essence, decreasing x leads to more smooth particles.

Table 3. the variation of the SI and mean coordination number of the particles with x. $r_1$ = 30-40, $r_2$ = 20-30, $r_3$ = 5-10, density = 0.65, RVE = $300^3$ voxels, number of particles = 88±3.

| x | SI | CN | mean particle radius |
|---|---|---|---|
| 10% | 0.59 ± 0.00 | 4.35 ± 0.02 | 34.89 ± 0.25 |
| 2% | 0.68 ± 0.00 | 3.84 ± 0.12 | 35.16 ± 0.56 |

The above results indicate that by tuning the input parameters, we can control the shape and surface roughness of the particles and thus produce random packings of non-spherical particles in a wide range of topologies, all while keeping the packing density and initial particle size distribution fixed.

## 2.5 Output

The output of the algorithm includes the particle coordination number spectrum, the number of inter-particle contacts, and a 3D image of the produced packing in Tagged Image File Format (TIFF) suitable for further analysis and simulation. Also, it is possible to produce 3D images with labeled particles (see Figure 7).

The algorithm is implemented in Python and C++ and is available in the Supplementary materials section.

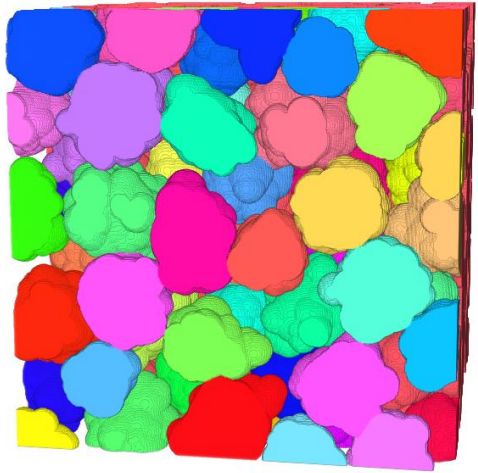

Figure 7. A random packing (with labeled particles) produced with inputs $30<r_1<40$, $15<r_2<20$, $10<r_3<15$, density=0.65, RVE size=$400^3$ voxels, number of particles = 190. Radii are chosen randomly from a uniform distribution.

## 2.6 Performance

To improve the performance of the algorithm in clustering the spheres, we only work with the nearest neighbors of a newly-added sphere. In this regard, we apply the concept of spatial indexing, specifically R*-trees, to index the spheres and to expedite their query. For this purpose, we make use of the libspatialindex library implemented using C++ [18]. Also, we store the 3D images in memory using C++ arrays to expedite random access to voxels during sphere clustering. Also, the location of the sphere surface voxels relative to the sphere center is calculated using the algorithm proposed by Biswas and Bhowmick [19]. Their algorithm utilizes the inherent symmetry of spheres to efficiently calculate the position of surface voxels using only primitive integer operations.

Typically, a $1000^3$-voxels RVE composed of 7000 particles in the radius range of 20-30 voxels with more than 300000 spheres clustered in total to reach a 60% density takes around 15 minutes to be produced using our algorithm on an average personal computer.

### 2.6.1 Random sequential addition (RSA)

One inherent limitation of RSA algorithms is their so-called 'jamming' limit [20], i.e., there is a cap on the number of spheres that can be randomly seeded in an RVE. In this regard, the number of failed attempts that are allowed during core seeding in the first stage, before moving on to the second stage, is an important value. This number dictates the duration of the first stage, and thus the final number of particles in the packing (see Figure 8); we call this number the 'escape factor.' To be able to produce uniform packings for all ranges of particle sizes, this factor should be tuned according to the fraction, $\text{mean core radius}/\text{RVE size}$, e.g., as the fraction decreases, the escape factor should be increased because of the high probability of sphere overlap. However, requiring a preset escape factor will unnecessarily encumber the user.

To obviate the need for a preset escape factor, we split each RVE into several sub-regions and carry out the random seeding process in each of these sub-regions independently. Not only this approach allows us to define a single escape factor based on the size of the sub-regions, but it also has the bonus of improving the efficiency of the RSA algorithm (See Figure 8). The suitable size of the sub-regions in the RVE can be chosen based on diagrams such as Figure 8.

In addition, as evident in Figure 8, the number of seeded spheres reaches a plateau in long durations. This observation denotes the mentioned 'jamming' limit. This limitation is mitigated by the fact that we form particles in three stages. Nevertheless, reproducing high-density packings of near-perfect spherical particles is not feasible using this algorithm.

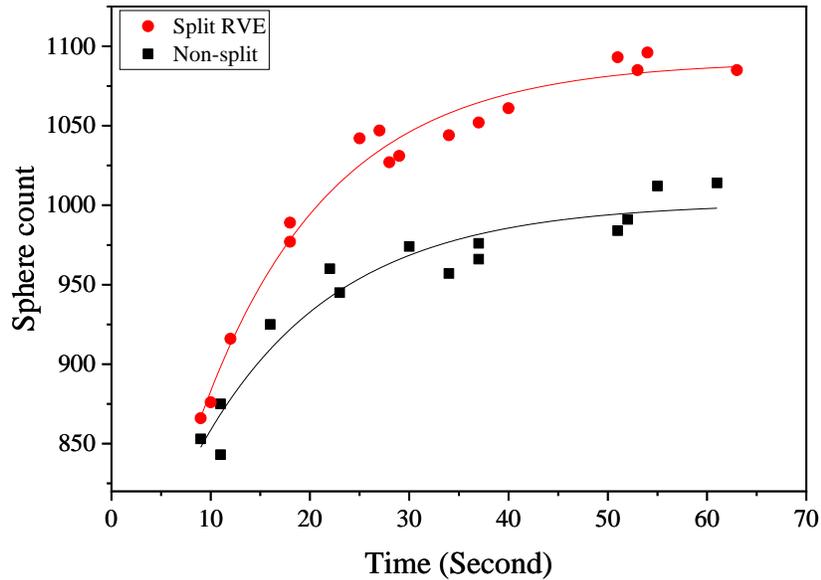

Figure 8. The increase of seeded spheres with time in the first stage for split and not split RVEs. The original RVE size is $500^3$ voxels, and the sub-regions are $100^3$ voxels. The radii of the core spheres lie in the range 20-30.

**2.6.2 Termination criterion**

In our algorithm, density is used as the termination criterion, i.e., sphere clustering is ended when the density of the packings reaches a predefined value. However, it should be noted that density can not be modified without affecting the connectivity of the particles. Table 4 illustrates the interconnectedness of density and connectivity in generated packings. As evident, increasing the density, with all other inputs fixed, leads to an increase in CN.

To explain the reason for this confoundment of packing density and connectivity, we should point that the density at the end of the core seeding stage (first stage) is dependent only on the core sphere radii. Therefore, to produce packings with different densities but similar particle size distributions (similar core sphere radii), varying numbers of secondary and tertiary spheres need to be deposited on the core spheres. This difference in the number of deposited spheres affects not only the shape and roughness of the particles but also the mean size of the particles. As seen in Table 4, the mean particle size increases slightly with density.

Though the increase of connectivity with density is intuitively expected, to be able to control the density, connectivity, and sphericity independently, we have to accommodate other termination criteria than density as well. Since our algorithm stores the voxels of the in-progress particles in memory during sphere clustering, it is possible to customize the algorithm so that the sphericity, size distribution, or connectivity of the particles can also be implemented as the termination criterion.

Table 4. the variation of the SI and mean coordination number of the particles with density. $r_1 = 20\text{-}30$, $r_2 = 15\text{-}20$, $r_3 = 10\text{-}15$, $X = 10\%$, RVE = $300^3$ voxels, number of particles = $230 \pm 5$.

| $\rho$ | SI | CN | mean particle radius |
|---|---|---|---|
| 0.55 | 0.76 ± 0.01 | 2.94 ± 0.17 | 24.25 ± 0.16 |
| 0.65 | 0.72 ± 0.00 | 4.29 ± 0.09 | 25.73 ± 0.43 |
| 0.75 | 0.70 ± 0.00 | 5.44 ± 0.15 | 26.66 ± 0.12 |

## 3. Euler characteristic and connectivity

The Euler characteristic $\chi$ and genus parameter $g$, two well-known mathematical quantities describing surface topology, are useful in monitoring the topological evolution of pore structures in sintered porous parts [10, 11, 21]. According to the theory of Gauss-Bonnet [21], Euler characteristic $\chi$ is calculated from the integral of Gaussian curvature $K$ over surface $A$ by Equation (5):

$$\chi = \frac{\pi}{2} \int_A K dA \tag{5}$$

Genus g is an integer representing the number of holes in a connected, orientable surface. For a single pore surface, Euler characteristic $\chi$ is mathematically related to Genus g according to Equation (6), and for N distinct pores, according to Equation (7).

$$\chi = 2 - 2g \qquad (6)$$

$$\chi/2 = \sum_{n=1}^{N}(1 - g) \qquad (7)$$

Since the sum of Genus is approximately equal to the number of holes $G$ in the pore surface, Equation (7) becomes $\chi/2 \approx N - G$, and for an interconnected pore structure, where $N=1$, we have:

$$\chi/2 \approx -G \qquad (8)$$

Based on equation (8), the variation of the Euler characteristic in a pore structure reflects the development of the holes in the pore surface. When monitored for the evolving pore structures of sintered specimens, the emergence and elimination of these holes reflect the formation of particle-particle contacts; this is illustrated in Figure 2 of our previous work [11].

On this note, for regular packings of spherical particles, Euler characteristic can be readily used to extract information about the number of particles, particle coordination number, etc.- even for random packings of spherical particles, extracting such information from the Euler characteristic has been done using numerically generated random packings [22]. However, for random packings of non-spherical particles the connection between the packing topology and Euler characteristic is less clear.

In this regard, the algorithm we have proposed allows us to produce a variety of random packings with known inter-particle connectivities that can be used to study the evolution of the Euler characteristic for packings of non-spherical particles. Figure 9 depicts the Euler characteristic as a function of the number of inter-particle contacts for packings produced by our algorithm. As evident, an increase in the Euler characteristic indicates a virtually equal rise in the number of inter-particle contacts. This observation denotes that the number of contacts is approximately equal to the number of holes ($\approx G$) in the interconnected pore structure of a random packing of non-spherical particles. This relationship is valuable since by monitoring the Euler characteristic and genus in real 3D reconstructions of pore structures, we can shed light on the state of the inter-particle contacts in sintered porous parts. This provides a significant tool in the study of the sintering process because, due to advances in 3D reconstruction techniques, capturing the evolution of real 3D pore structures during sintering is now accessible and straightforward and by means of Euler characteristic so is tracking the formation and merging of particle-particle contacts. As discussed earlier, the evolution of inter-particle contacts plays a significant role in the sintering kinetics and the development of macroscopic properties in porous sintered parts.

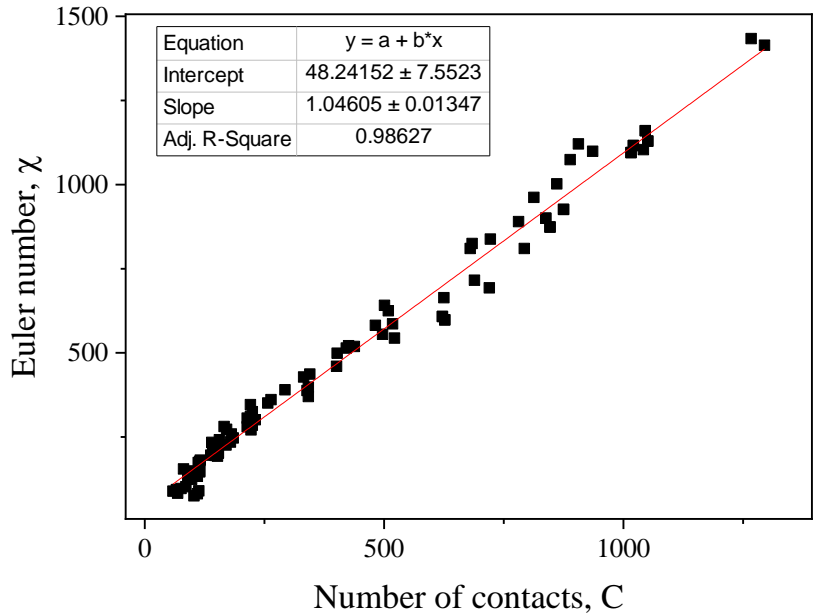

Figure 9. Variation of the Euler characteristic with the number of inter-particle contacts in random packings of non-spherical particles.

## 4. Conclusion

We introduced an algorithm that can be used to produce random packings of non-spherical particles. To create non-spherical particles, we cluster three groups of spheres, each in a different size range, in three different stages. The use of varying-sized spheres allows us to tune the shape and surface roughness of the particles and generate packings in a wide range of topologies, which can be used to study the role of inter-particle contacts in sintering-based processes.

Another possible use of our algorithm is to capture the topology of the real pore structure of sintered porous parts by tuning the inputs of the algorithm to mimic the Euler characteristic of the real structure. Also, with additive manufacturing techniques, it is possible to produce prototypes with exact connectivity to scrutinize the role of interfacial regions in the development of macroscopic properties, such as permeability and mechanical strength, in sintered porous parts.